\pdfoutput=1

\documentclass[11pt,twoside,a4paper,cmspaper,final,collab]{cms-tdr}
\def\svnVersion{0a3016f-D}\def\svnDate{2019/08/07}\def\cmsCernNoTag{CERN-EP-2018-257}\def\cmsCernDate{\today}\def\cmsMessage{"Published in Physics Letters B as \href{http://dx.doi.org/10.1016/j.physletb.2019.07.014}{\doi{10.1016/j.physletb.2019.07.014}.}"}

\begin{document}\cmsNoteHeader{HIN-17-008}

\hyphenation{had-ron-i-za-tion}
\hyphenation{cal-or-i-me-ter}
\hyphenation{de-vices}
\RCS$HeadURL$
\RCS$Id$
\newlength\cmsFigWidth
\ifthenelse{\boolean{cms@external}}{\setlength\cmsFigWidth{0.98\columnwidth}}{\setlength\cmsFigWidth{0.69\textwidth}}
\ifthenelse{\boolean{cms@external}}{\providecommand{\cmsLeft}{top\xspace}}{\providecommand{\cmsLeft}{left\xspace}}
\ifthenelse{\boolean{cms@external}}{\providecommand{\cmsRight}{bottom\xspace}}{\providecommand{\cmsRight}{right\xspace}}
\newcommand{\sqrts}{ \ensuremath{\sqrt{s}}\xspace}

\newcommand{\RAA}{\ensuremath{R_{\mathrm{AA}}}\xspace}
\newcommand{\TAA}{\ensuremath{T_{\mathrm{AA}}}\xspace}
\newcommand{\PbPb}{\ensuremath{\mathrm{PbPb}}\xspace}
\newcommand{\pp}{\Pp{}\Pp}
\newlength\cmsTabSkip\setlength{\cmsTabSkip}{1ex}
\providecommand{\Pphi}{\ensuremath{\phi}}
\providecommand{\PABzs}{\ensuremath{\overline{\PBzs}}\xspace}
\providecommand{\Bzerosdecay}{\ensuremath{\PBzs \to \PJGy~\Pphi \to \Pgmp\Pgmm\PKp\PKm}\xspace}
\providecommand{\BzeroKstar}{\ensuremath{\PBz\to\PJGy~\PKst^0}\xspace}
\providecommand{\HYDJET}{\textsc{hydjet}\xspace}
\providecommand{\NA}{\ensuremath{\text{---}}}

\cmsNoteHeader{HIN-17-008}

\title{Measurement of $\PBzs$ meson production in \pp and \PbPb collisions at $\sqrtsNN = 5.02\TeV$}

\date{\today}

\abstract{
The production cross sections of $\PBzs$ mesons and charge conjugates are measured in proton-proton (\pp) and \PbPb collisions via the exclusive decay channel \Bzerosdecay at a center-of-mass energy of $5.02\TeV$ per nucleon pair and within the rapidity range $\abs{y}<2.4$ using the CMS detector at the LHC. The \pp measurement is performed as a function of transverse momentum (\pt) of the $\PBzs$ mesons in the range of 7 to 50\GeVc and is compared to the predictions of perturbative QCD calculations. The $\PBzs$ production yield in \PbPb collisions is measured in two \pt intervals, 7 to 15 and 15 to 50\GeVc, and compared to the yield in \pp collisions in the same kinematic region. The nuclear modification factor (\RAA) is found to be $1.5 \pm 0.6\stat\pm 0.5\syst$ for 7--15\GeVc, and $0.87 \pm 0.30\stat\pm 0.17\syst$ for 15--50\GeVc, respectively. Within current uncertainties, the $\PBzs$ results are consistent with models of strangeness enhancement, and suppression by parton energy loss, as observed for the $\PBp$ mesons.}

\hypersetup{%
pdfauthor={CMS Collaboration},%
pdftitle={Measurement of the Bs0 meson production in pp and  PbPb collisions at sqrt(s[NN]) = 5.02 TeV},%
pdfsubject={CMS},%
pdfkeywords={physics, quark gluon plasma, B meson, open heavy-flavor}}

\maketitle

\section{Introduction}

Relativistic heavy ion collisions allow the study of quantum chromodynamics (QCD) at high energy density and temperature. Under such extreme conditions, a state consisting of deconfined quarks and gluons, the quark-gluon plasma (QGP)~\cite{QGP1,QGP2}, is predicted by lattice QCD calculations~\cite{Karsch:2003jg}.
The study of the phenomenon in which the outgoing partons interact strongly with the QGP and lose energy by means of elastic collisions and medium-induced gluon radiation~\cite{Eloss1,Baier:2000mf,Chatrchyan:2011sx,Aad:2010bu,Andronic:2015wma} can provide insights into the energy density and diffusion properties of the QGP. Heavy quarks are effective probes to study these properties of the medium. Charm and beauty quarks that are primarily produced in hard scatterings at the early stages of the collision are expected to carry the full evolution history of the QGP formation~\cite{Andronic:2015wma}. On the other hand it is expected~\cite{PhysRevLett.48.1066} that, via the process $\Pg\Pg \to \cPqs\cPaqs$, an enhancement of strangeness in a thermally and chemically equilibrated QGP should occur if its temperature is above the strange quark mass. Measurements at the BNL RHIC of the production of strange baryons and mesons, using different collision systems and beam energies, provide systematic support for this expectation~\cite{Agakishiev:2011ar,Abelev:2008zk,Arsene:2009jg,Adamczyk:2017wsl,Adare:2015ema}. Because of the interplay between the predicted enhancement of strange quark production and the quenching mechanism of beauty quarks, the measurement of strange beauty particles is important for studying the mechanisms of beauty hadronization in heavy ion collisions. In the presence of a medium with increased strangeness content~\cite{ALICE:2017jyt,Abelev:2013xaa}, the relative yield of $\PBzs$ mesons with respect to nonstrange beauty mesons at transverse momentum (\pt) below $\sim$15\GeVc~\cite{Andronic:2015wma,He:2014cla} can be enhanced in nucleus-nucleus collisions compared to proton-proton (\pp) interactions. This can happen if recombination is a significant factor of beauty hadronization in the QGP~\cite{Molnar:2003ff,Greco:2003mm,Greco:2003vf}. The recombination processes, which are considered markers for the presence of a deconfined medium, were most recently tested in the open charm sector by the ALICE Collaboration~\cite{Acharya:2018hre}. A possible hint for an enhancement in the relative yield of \PsDp\ mesons with respect to nonstrange charmed mesons for $\pt < 8$\GeVc in central \PbPb collisions at a center-of-mass energy of $\sqrtsNN=5.02\TeV$ per nucleon pair was observed.

The production of $\PBzs$ mesons was previously measured at the CERN LHC by the CMS Collaboration in \pp collisions at a center-of-mass energy of $\sqrts=7\TeV$~\cite{Chatrchyan:2011vh} and in proton-lead (pPb) collisions at $\sqrtsNN=5.02\TeV$~\cite{Khachatryan:2015uja}. In this letter, we report the first measurement of exclusive $\PBzs$ meson decays ever performed in nucleus-nucleus collisions and in \pp collisions at 5.02\TeV. The \pp measurement is performed as a function of \pt and compared to the predictions of fixed-order plus next-to-leading order logarithmic (FONLL) perturbative QCD calculations~\cite{FONLLcharmbottomPP1,FONLLcharmbottomPP2,FONLLcharmbottomPP3}. The nuclear modification factor (\RAA) of $\PBzs$ mesons, which is defined as the ratio of the yield in PbPb collisions with respect to that in pp collisions scaled by the corresponding number of binary nucleon-nucleon (NN) collisions, is shown. The comparison between the \RAA of $\PBzs$ mesons and that of $\PBp$ mesons measured by CMS at the same energy~\cite{Sirunyan:2017oug} is also presented.

The $\PBzs$ meson and its charge conjugate are measured in the rapidity range $\abs{y}<2.4$ via the reconstruction of the decay channel \Bzerosdecay, which has the branching fraction $\mathcal{B} = (3.12 \pm 0.24) \times 10^{-5}$~\cite{Olive:2016xmw}. The \pp measurement is performed as a function of the $\PBzs$ \pt in three intervals, 7--15, 15--20, and 20--50\GeVc. The \PbPb production yield and the \RAA measurement are performed in two \pt intervals, 7--15 and 15--50\GeVc, inclusively for all events (\ie, 0--100\% centrality, the degree of overlap of the two colliding nuclei). Throughout the letter, unless otherwise specified, the $y$ and \pt variables given are those of the $\PBzs$ mesons. This analysis does not distinguish between the charge conjugates.

\section{Experimental apparatus and data sample}

The central feature of the CMS detector is a superconducting solenoid, which provides a magnetic field of 3.8\unit{T}. Within the solenoid volume are a silicon tracker that measures charged particles in the pseudorapidity range $\abs{\eta} < 2.5$, a lead tungstate crystal electromagnetic calorimeter, and a brass and scintillator hadron calorimeter. For charged particles of $1 < \pt < 10$\GeVc and $\abs{\eta} < 1.4$, the track resolutions are typically 1.5\% in \pt and 25--90 (45--150)\mum in the transverse (longitudinal) impact parameter \cite{TRK-11-001}. Muons are measured in the range $\abs{\eta} < 2.4$, with detection planes made using three technologies: drift tubes, cathode strip chambers, and resistive-plate chambers. The muon reconstruction algorithm starts by finding tracks in the muon detectors, which are then fitted together with tracks reconstructed in the silicon tracker to form "global muons". Matching muons to tracks measured in the silicon tracker results in a relative \pt resolution for muons with $20 < \pt < 100$\GeVc of 1.3--2.0\% in the barrel ($\abs{\eta} < 1.2$) and better than 6\% in the endcaps ($1.6 < \abs{\eta} < 2.4$). For muons with higher \pt up to 1\TeVc, the \pt resolution in the barrel is better than 10\%~\cite{Chatrchyan:2012xi}.  The hadron forward (HF) calorimeter uses steel as an absorber and quartz fibers as the sensitive material. The two halves of the HF are located 11.2\unit{m} away from the interaction point, one on each end, providing together coverage in the range $3.0 < \abs{\eta} < 5.2$. In this analysis, the HF information is used for performing an offline event selection. A detailed description of the CMS experiment and coordinate system can be found in Ref.~\cite{bib_CMS}.

Several Monte Carlo (MC) simulated event samples are used to evaluate background components, signal efficiencies, and detector acceptance corrections. The simulations include samples containing only the $\PBzs$ meson decay channels being measured, and samples with inclusive (prompt and nonprompt) $\PJGy$ mesons. Proton-proton collisions are generated with {\PYTHIA}8 v212~\cite{Sjostrand:2014zea} tune CUETP8M1~\cite{Khachatryan:2015pea} and propagated through the CMS detector using the \GEANTfour package~\cite{Allison:2016lfl}. The decay of the $\PBzs$~mesons is modeled with \EVTGEN 1.3.0~\cite{Lange:2001uf}, and final-state photon radiation in the $\PBzs$ decays is simulated with \PHOTOS 2.0~\cite{Barberio:1990ms}. For the $\PbPb$ MC samples, each {\PYTHIA}8 event is embedded into a $\PbPb$ collision event generated with {\HYDJET}~1.8~\cite{Lokhtin:2005px}, which is tuned to reproduce global event properties, such as the charged-hadron \pt spectrum and particle multiplicity. For both samples, the signal \pt shape is reweighted to match the one from FONLL. For both \pp and \PbPb data and MC samples, the dimuon and ditrack mass distributions/resolutions are consistent. 

Events were collected with the same trigger during the \pp and \PbPb data  acquisition, requiring the presence of two muon candidates (with no explicit momentum threshold) in coincidence with a bunch crossing. For the offline analysis, events have to pass a set of selection criteria designed to reject events from background processes (beam-gas collisions and beam scraping events) as described in Ref.~\cite{Khachatryan:2016odn}. Events are required to have at least one reconstructed primary interaction vertex, formed by two or more tracks, with a distance from the center of the nominal interaction region of less than 15\cm along the beam axis. In \PbPb collisions, the shapes of the clusters in the pixel detector have to be compatible with those expected from particles produced by a \PbPb collision~\cite{Khachatryan:2010xs}. In order to select hadronic collisions, the \PbPb events are also required to have at least three towers in each of the HF detectors with energy deposits of more than 3\GeV per tower. The combined efficiency for this event selection, including the remaining non-hadronic contamination, is $(99\pm2)$\%. Values higher than 100\% are possible, reflecting the potential presence of ultra-peripheral (i.e., non-hadronic) collisions in the selected event sample. The \PbPb sample corresponds to an integrated luminosity of approximately 351\mubinv. This value is indicative only, as the \PbPb yield is normalized by the total number of minimum bias events sampled, $N_{\text{MB}}$~\cite{Khachatryan:2016odn}. The \pp data set corresponds to an integrated luminosity of 28.0\pbinv, which is known to an accuracy of $\pm$2.3\% from the uncertainty in the calibration based on a van der Meer scan~\cite{CMS-PAS-LUM-16-001}. The average number of additional collisions per bunch crossing is approximately 0.9 for \pp and less than 0.01 for \PbPb data. The presence of multiple collisions is found to have a negligible effect on the measurement.

\section{Signal extraction}

The analysis procedure is common for \pp and \PbPb data. Kinematic limits are imposed on the single muons so that their reconstruction efficiency stays above 10\%. These limits are $\pt^{\mu}>3.5\GeVc\enspace \text{ for } \abs{\eta^{\mu}}<1.2$, $\pt^{\mu}>1.8\GeVc\text{ for }2.1\le\abs{\eta^{\mu}}<2.4$, and linearly interpolated in the $1.2<\abs{\eta^{\mu}}<2.1$ region. The muons are also required to match the muons that triggered the event online, and to pass selection criteria optimized for low \pt (the so-called \textit{soft selection}~\cite{Chatrchyan:2012xi}). Two muons of opposite sign (OS), with an invariant mass within $\pm$150\MeVcc of the world-average $\PJGy$ meson mass~\cite{Olive:2016xmw} are selected to reconstruct a $\PJGy$ candidate, with a mass resolution of typically 18--55\MeVcc, depending on the dimuon rapidity and \pt. The OS muon pairs are fitted with a common vertex constraint and are kept if the p-value of the $\chi^2$ of the fit is greater than 1\%, thus lowering the background from charm and beauty hadron semileptonic decays. Similarly, the $\phi$ meson candidates are formed with a common vertex constraint between two OS charged-particle tracks with $\pt>300 (150)$\MeVc for \PbPb (\Pp\Pp) sample, both required to pass standard selections~\cite{Khachatryan:2016odn}. The invariant mass, with a resolution of $\sim$3.9 (3.4)\MeVcc for \PbPb (\Pp\Pp) data, is required to be within 15\MeVcc of the world-average $\phi$ meson mass~\cite{Olive:2016xmw}. The $\PBzs$~meson candidates are constructed by combining the $\PJGy$ and $\phi$ candidates and requiring that they originate from a common vertex. Without using particle identification, assumptions need to be made about the masses of the charged particles. The difference between the natural width (according to PDG~\cite{PhysRevD.98.030001}) and the measured width (reflecting detector resolution) of the peaks is much bigger for the $\PJGy$ meson than for the $\phi$ meson. Therefore, in calculating the mass of the $\PBzs$ candidates, the two charged particles are always assumed to have the mass of charged kaons, and the muon pair is assumed to have the mass of a $\PJGy$ meson. 

The $\PBzs$ candidates are selected according to their daughter charged particle track kinematics, the $\chi^{2}$ probability of their decay vertex (the probability for the muon tracks from the $\PJGy$ meson decay and the other charged particle tracks to originate from a common vertex), the distance between the primary and decay vertices (normalized by its uncertainty), and the pointing angle (the angle between the line segment connecting the primary and decay vertices and the momentum vector of the $\PBzs$ meson). The selection is optimized separately for \pp and \PbPb results as well as each individual \pt bin, using a multivariate technique that employs the boosted decision tree (BDT) algorithm~\cite{Hocker:2007ht}, in order to maximize the statistical significance of the $\PBzs$ meson signals. The $\PBzs$ signal samples are taken from simulation. The signal samples are scaled to the number of $\PBzs$ candidates predicted by FONLL calculations corresponding to the integrated luminosity of the analyzed data sample. This normalization is not used when performing the BDT training. The background samples for the multivariate training are taken from data sidebands of the $\PBzs$ meson invariant mass ($0.2<\abs{M_{\mu\mu\PK\PK}-M_{\PBzs, {\mathrm{PDG}}}}<0.3\GeVcc$), which is about 5$\sigma$ away from the PDG \PBzs mass value. The optimal selection criterion is the working point with the highest signal significance ($N_{\mathrm{s}}/\sqrt{(N_{\mathrm{s}}+N_{\mathrm{b}}}$), where $N_{\mathrm{s}}$ ($N_{\mathrm{b}}$) are the expected signal (background) candidate yields from the simulated signal (data sidebands) within the mass range $\abs{M_{\mu\mu\PK\PK}-M_{\PBzs, {\mathrm{PDG}}}}<0.08\GeVcc$.

\begin{figure*}[tb]
\centering
\includegraphics[width=.49\textwidth]{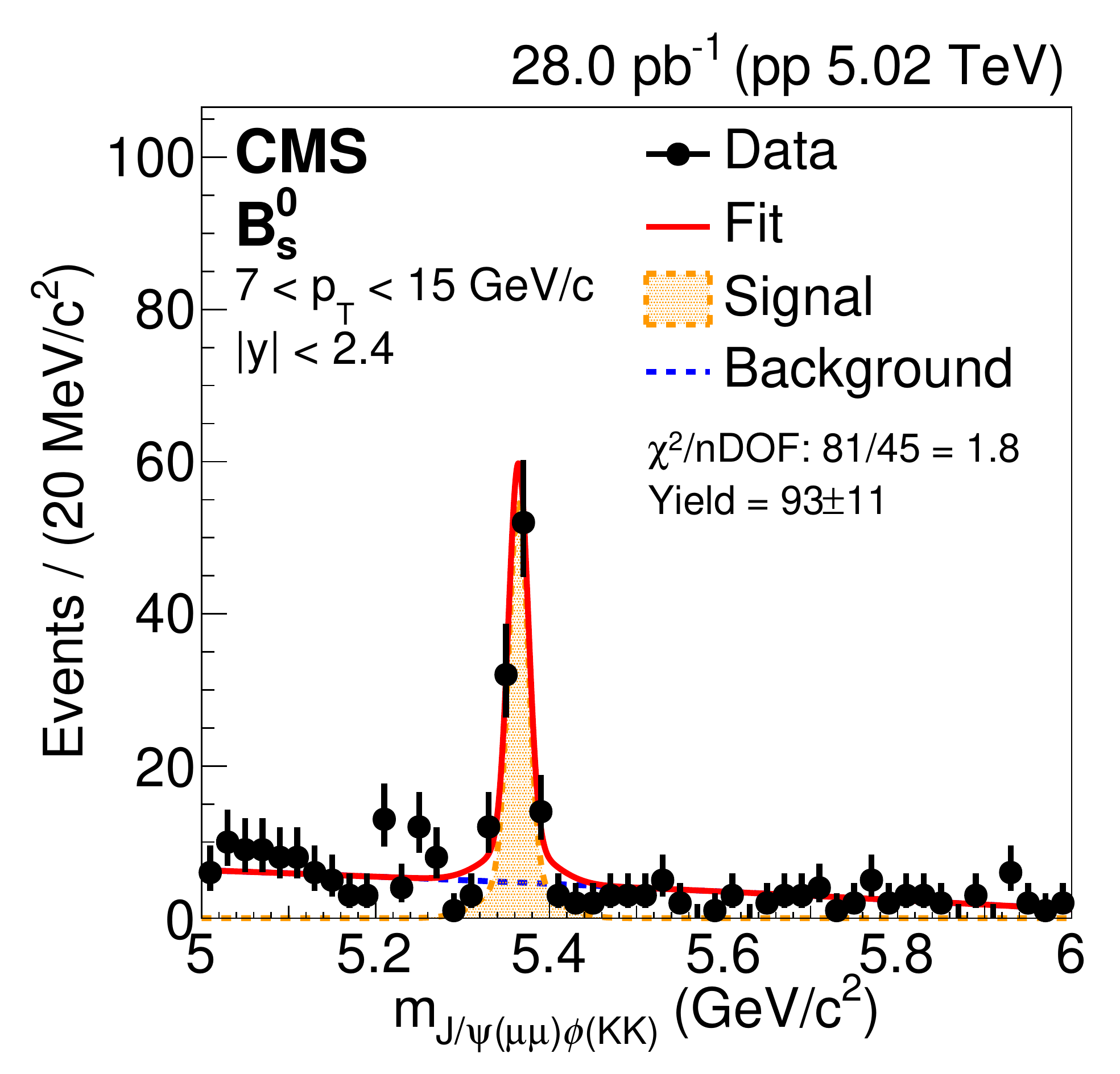}
\includegraphics[width=.49\textwidth]{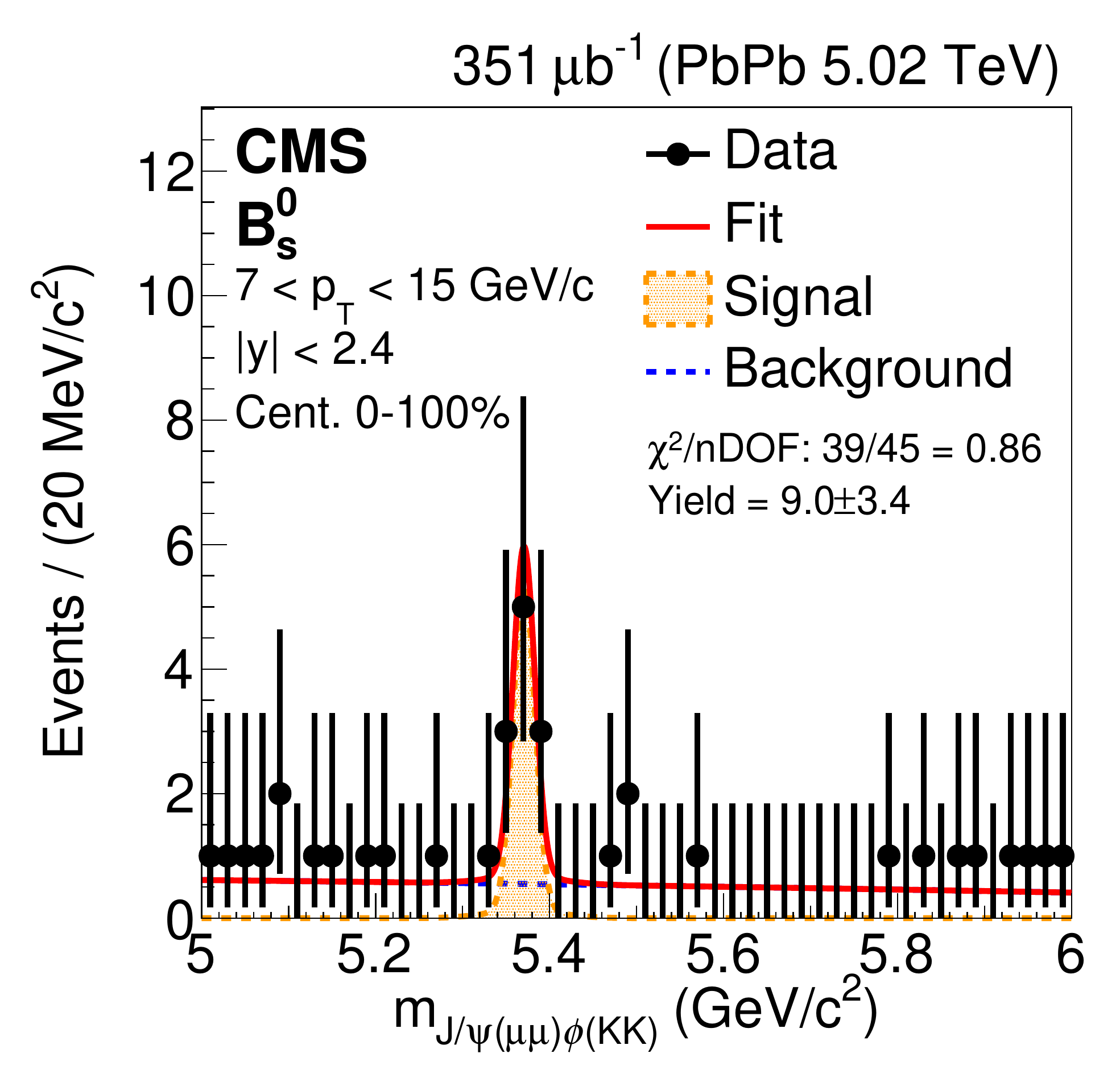}
\caption{Invariant mass distributions of $\PBzs$ candidates in \pp (\cmsLeft) and \PbPb (\cmsRight) collisions measured in the range $\abs{y}<2.4$ and in the \pt range of 7--15\GeVc. The $\chi^2$ divided by the number of degrees of freedom (nDOF) is also given.}
\label{fig:rawYieldsBsmeson1}
\end{figure*}

\begin{figure*}[tb]
\centering
\includegraphics[width=.49\textwidth]{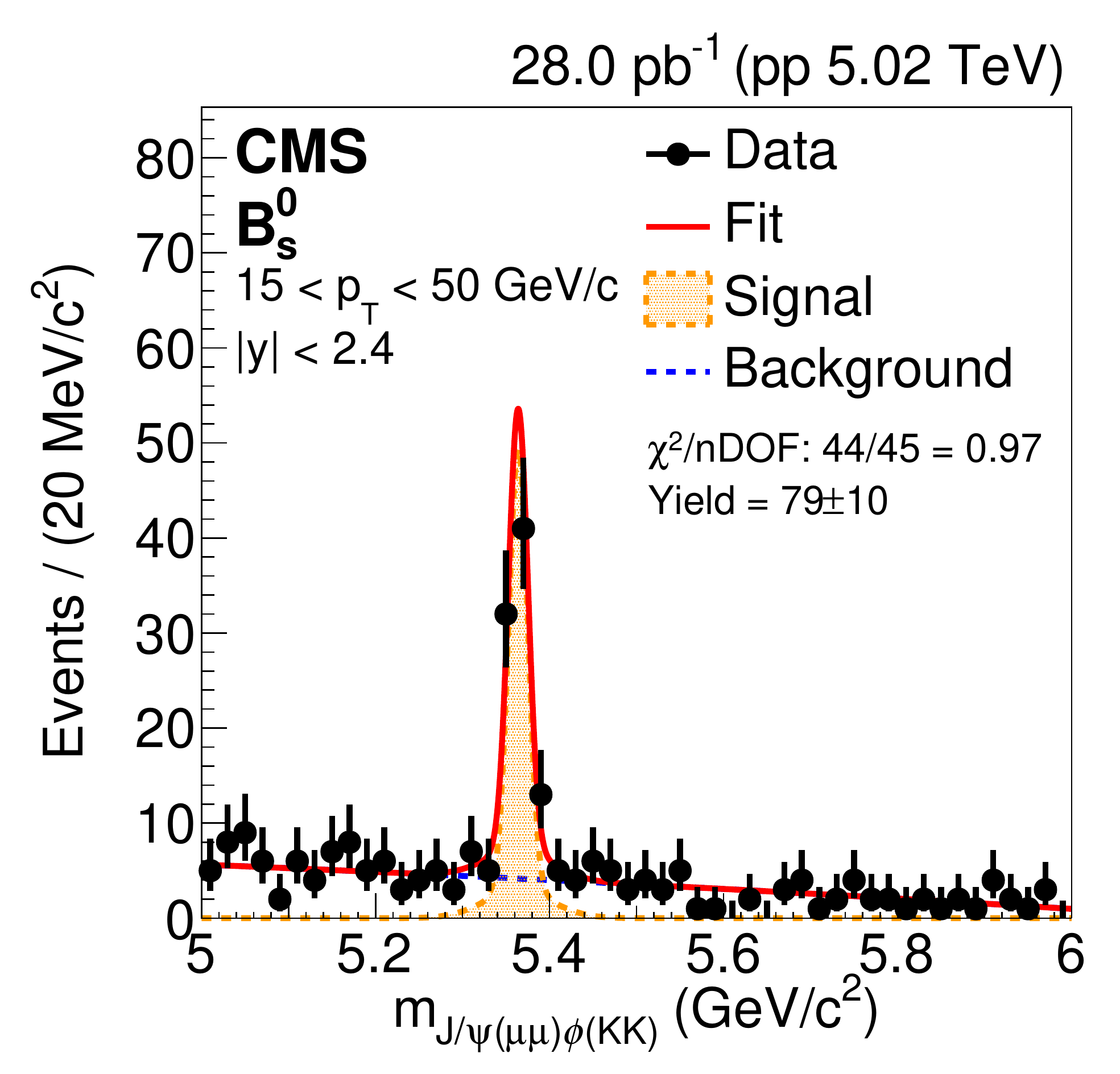}
\includegraphics[width=.49\textwidth]{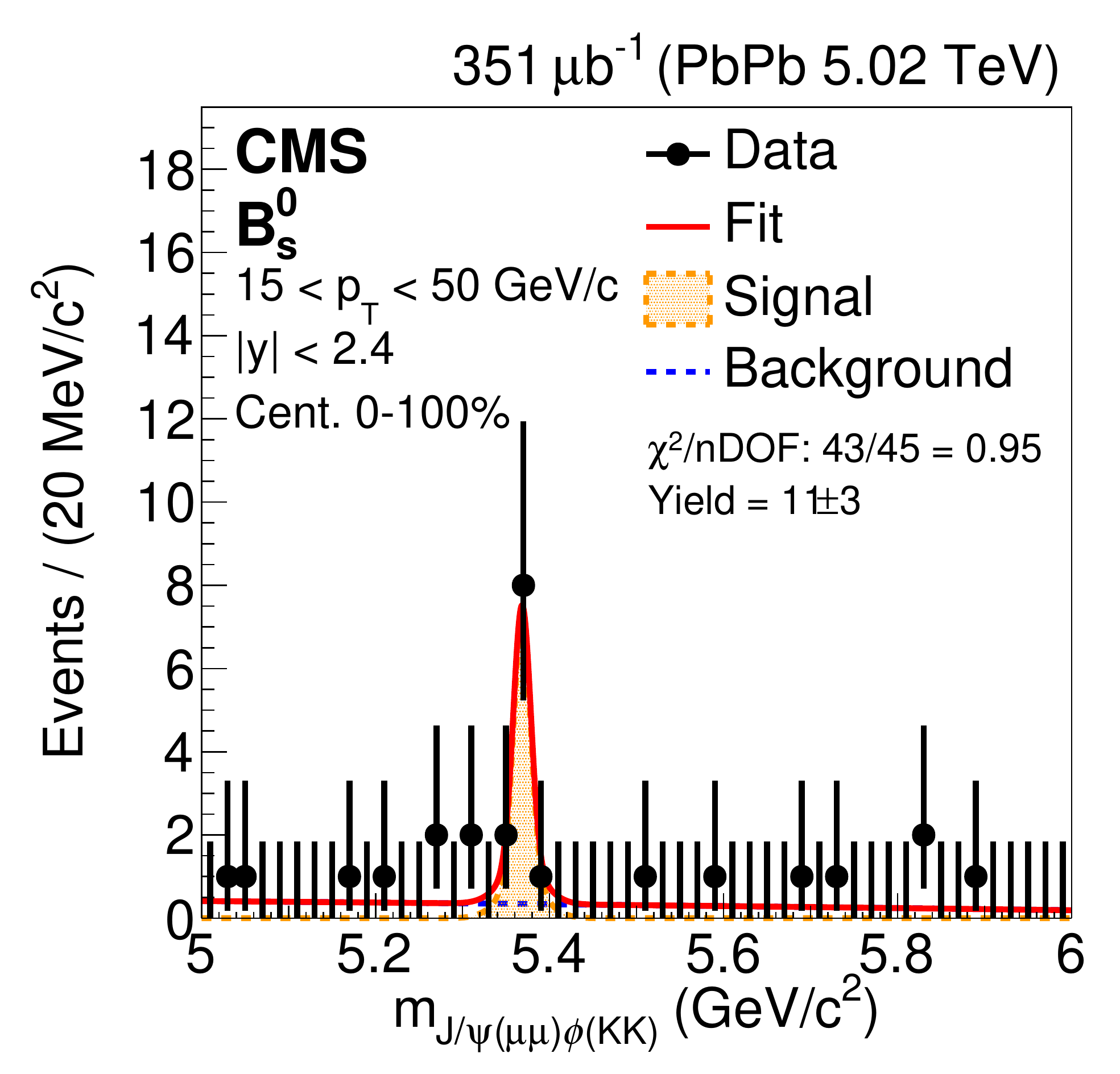}
\caption{Invariant mass distributions of $\PBzs$ candidates in \pp (\cmsLeft) and \PbPb (\cmsRight) collisions measured in the range $\abs{y}<2.4$ and in the \pt range of 15--50\GeVc. The $\chi^2$ divided by the number of degrees of freedom (nDOF) is also given.}
\label{fig:rawYieldsBsmeson2}
\end{figure*}

The raw yields of \PBzs\ mesons in \pp and \PbPb collisions are extracted using an extended unbinned maximum likelihood fit to the invariant mass distribution of the \PBzs\  candidates in the mass range 5--6$\GeVcc$. The estimation of the statistical uncertainties of the fitted raw yields is based on the second derivatives of the negative log-likelihood function. Examples of fits to the invariant mass distributions in \pp and \PbPb collisions are
shown in Figs.~\ref{fig:rawYieldsBsmeson1} and \ref{fig:rawYieldsBsmeson2} for the $\pt$ regions 7--15 and 15--50\GeVc, respectively. The signal shape is modeled by two Gaussian functions with a common mean (which is a free parameter together with the amplitude), and different widths individually determined from MC simulations for the \pp and \PbPb results. The relative contribution of the two Gaussian functions to the signal yield is also fixed at the value given by the MC sample. The background is dominated by random combinations of prompt and nonprompt \PJGy\ candidates with extra particles and it is modeled by a first-order polynomial, as determined by studies of the inclusive $\PJGy$ MC sample. Peaking structures that could arise from the background contamination of other \PB\ meson decays (e.g., \BzeroKstar) were found to be negligible as a consequence of the tight selection on the mass of the $\phi$ candidate.

The differential cross section for $\PBzs$ production in $\abs{y}<2.4$ is computed in each \pt interval according to
\begin{linenomath}
\begin{equation}
 \left.\frac{{\rd}\sigma^{\PBzs}}{{\rd}\pt}\right|_{\abs{y}<2.4} =
 \frac{1}{2} \frac{1}{\mathcal{B} \, \mathcal{L}} \frac{1}{\Delta\pt} \left.\frac{N^{(\PBzs + \PABzs)}_{\pp}(\pt)}{\alpha_{{\Pp\Pp}}(\pt) \, \epsilon_{\pp}(\pt)}\right|_{\abs{y}<2.4},
 \label{eq:crosssectionDpPb}
\end{equation}
\end{linenomath}
for \pp data, and for \PbPb data according to
\ifthenelse{\boolean{cms@external}}{
\begin{multline}
 \left. \frac{1}{\TAA}\frac{{\rd}N^{\PBzs}_\PbPb}{{\rd}\pt}\right|_{\abs{y}<2.4} = \frac{1}{2} \frac{1}{\mathcal{B} \, N_{\text{MB}} \, \TAA} \frac{1}{\Delta\pt} \\
 \times\left.\frac{N^{(\PBzs + \PABzs)}_\PbPb(\pt)}{\alpha_\PbPb(\pt) \, \epsilon_\PbPb(\pt)}\right|_{\abs{y}<2.4} .
 \label{eq:PbPbYield}
\end{multline}
}{
\begin{linenomath}
\begin{equation}
 \left. \frac{1}{\TAA}\frac{{\rd}N^{\PBzs}_\PbPb}{{\rd}\pt}\right|_{\abs{y}<2.4} = \frac{1}{2} \frac{1}{\mathcal{B} \, N_{\text{MB}} \, \TAA} \frac{1}{\Delta\pt} \left.\frac{N^{(\PBzs+\PABzs)}_\PbPb(\pt)}{\alpha_\PbPb(\pt) \, \epsilon_\PbPb(\pt)}\right|_{\abs{y}<2.4} .
 \label{eq:PbPbYield}
\end{equation}
\end{linenomath}
}
The $N^{(\PBzs + \PABzs)}_{\pp, \PbPb}$ is the raw signal yield extracted in each \pt interval of width $\Delta \pt$, $(\alpha, \epsilon)_{\pp, \PbPb}$ represents the corresponding acceptance times efficiency, and $\mathcal{B}$ is the branching fraction of the decay chain. For the \pp cross section, $\mathcal{L}$ represents the integrated luminosity, and for the \PbPb cross section,  $N_{\text{MB}}$ is the number of minimum bias events and \TAA is the nuclear overlap function~\cite{Miller:2007ri}. The \TAA is equal to the number of NN binary collisions divided by the NN total inelastic cross section, and it can be interpreted as the NN-equivalent integrated luminosity per heavy ion collision. The \TAA value for inclusive \PbPb collisions at $\sqrtsNN=5.02$\TeV is $(5.6\pm0.2)$\mbinv as estimated from an MC Glauber model~\cite{Miller:2007ri, Khachatryan:2016odn}. Assuming that, in the kinematic region accessible by the present measurement, the $\PBzs$ and $\PABzs$ production cross sections are equal, the factor 1/2 accounts for the fact that the yields are measured for particles and antiparticles added together, but the cross section is given for one species only.

\begin{figure}[tb]
\centering
\includegraphics[width=.45\textwidth]{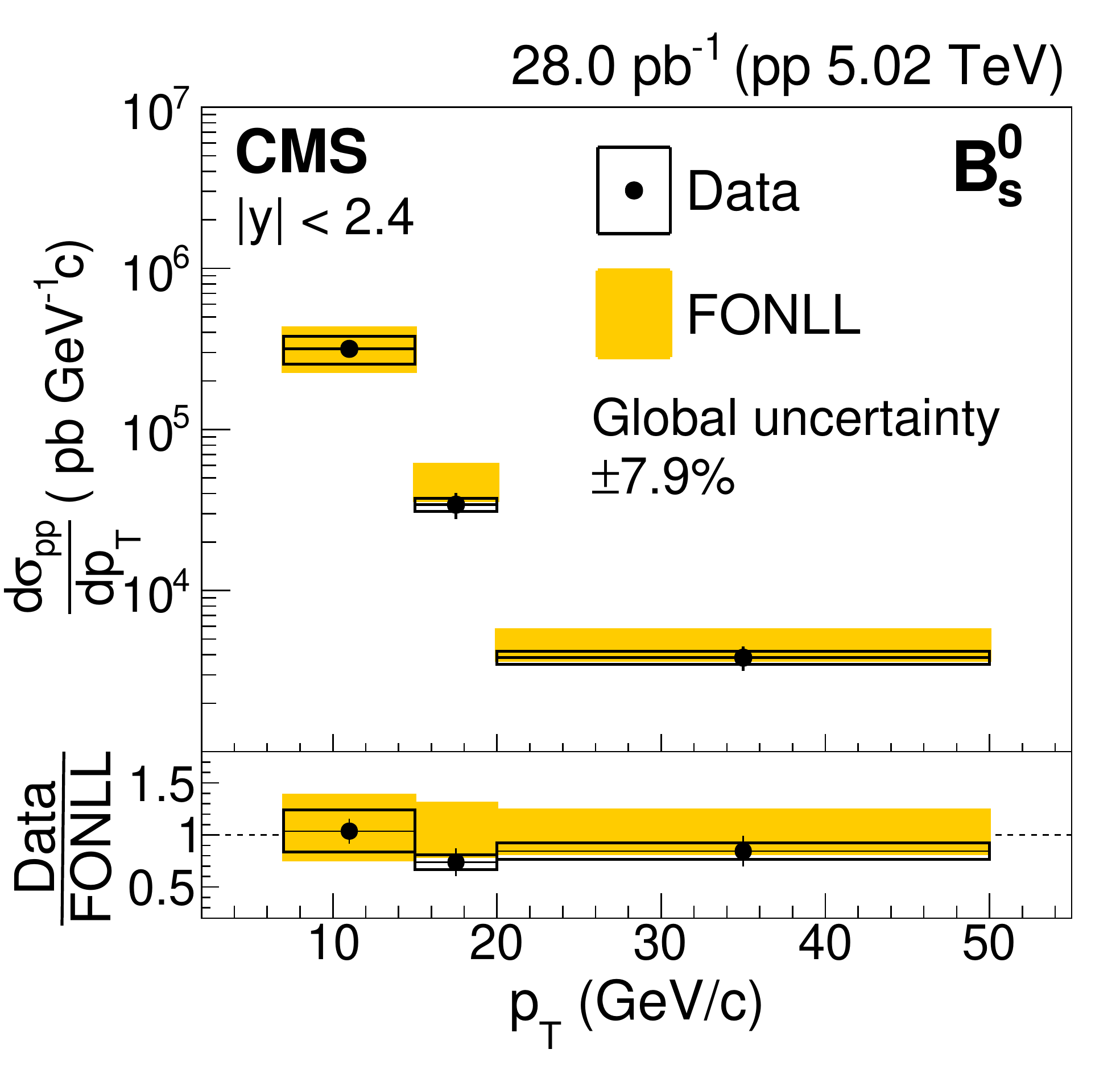}
\caption{
The \pt-differential production cross section of \PBzs\ in \pp collisions at $\sqrts=5.02\TeV$ in three \pt intervals from 7 to 50\GeVc.
The vertical bars (boxes) correspond to statistical (systematic) uncertainties. The global systematic uncertainty, listed in the legend and not included in the point-to-point uncertainties, comprises the uncertainties in the integrated luminosity measurement and in the branching fraction $\mathcal{B}$.
The \pp cross section is compared to FONLL calculations~\cite{FONLLcharmbottomPP3} represented by the colored yellow boxes with the heights indicating the theoretical uncertainty.}
\label{fig:crosssections}
\end{figure}

\section{Systematic uncertainties}

The cross section measurements are affected by several sources of systematic uncertainties arising from the signal extraction, corrections, $\mathcal{B}$, $\mathcal{L}$,  $N_{\text{MB}}$, and \TAA determination. Unless mentioned otherwise, the same procedures were used to estimate the uncertainties for the \pp and \PbPb results. The uncertainty of the signal modeling is evaluated by considering four fit variations: (i) increasing/decreasing the width parameters determined from simulation by 4\% (the maximum relative statistical uncertainty of the fitted width parameter among all \pt bins from \pp and \PbPb data); (ii) using a single Gaussian function; (iii) using a sum of three Gaussian functions with a common mean, and, (iv) fixing the mean of the Gaussian function to the value determined from simulation. The uncertainty in the modeling of the background shapes is also evaluated by varying the probability distribution functions used to describe the background to a higher-order polynomial and exponential function. The maximum of the signal variations and the maximum of all the background variations are  propagated as systematic uncertainties. For the \pp results, the systematic uncertainty due to the selection of the $\PBzs$~meson candidates is estimated by comparing the BDT-obtained nominal result with the results using a cut-based method (a rectangular cut) that uses the Genetic Algorithm to determine the best cut value for each parameter~\cite{Hocker:2007ht}. The same signal and background shape parametrization are used, and the same analysis parameters are optimized as in the BDT nominal method. The significance is similar for the two methods ($\sim$8) for the \pp bins. This provides an estimate of the potential difference between different selection criteria. The full difference between the two methods is propagated as a systematic uncertainty. For the \PbPb results, because of the small signal in data, in order to minimize the impact of statistical fluctuations, a different approach was taken. In this case, the $\PBzs$ selection uncertainty was estimated using the pp data sample, as the full difference in the yield between the \pp results with the BDT trained on the \pp sample (the nominal result) and the results with the BDT trained on the \PbPb sample (the selection used for the \PbPb results).

The bin-by-bin systematic uncertainties associated with the acceptance correction are estimated by varying the shape of the generated $\PBzs$~meson \pt and $y$ spectra. For the purpose of the systematic studies only, both data and MC are split into four \pt and $y$ bins. The ratio between data and simulated \pt spectra (including their statistical uncertainties) is used to generate pseudo-experiments (`toys'). Each toy is fit with a polynomial, which is then used to reweight the MC $\PBzs$~meson \pt spectra. A new acceptance value is calculated for each modified shape, for each kinematic bin. The root mean square (RMS) of all acceptances determined via toys is propagated as the systematic uncertainty by choosing the maximum RMS value emerging from the \pt and $y$ shape variations. Because of the small signal available, for the \PbPb results the \pp ratio is used to generate the toys. There is also an uncertainty assigned to account for potential bias in the efficiency calculations from the FONLL simulations of the $\PBzs$~meson \pt shape. This uncertainty is calculated as the difference between the nominal results and those obtained by generating the {\PYTHIA} \pt shape. An additional uncertainty comes from the finite size of the MC samples. This is determined by the statistical uncertainty of the simulated signal, after applying all selection criteria.

The uncertainty in the efficiency of the muon trigger, reconstruction, and identification is evaluated bin-by-bin using control samples in data~\cite{Khachatryan:2010xn}. A relative systematic uncertainty of 4\% per hadron track in \pp collisions~\cite{TRK-11-001} and 6\% in \PbPb collisions~\cite{Khachatryan:2016odn} is also considered, to account for the uncertainty in the track reconstruction efficiency. This uncertainty propagates to 8\% and 12\% for the $\PBzs$ measurement in \pp and \PbPb, respectively. The systematic uncertainty in the cross section measurement is computed as the sum in quadrature of the different contributions mentioned above. The uncertainty in the $\PBzs$~meson decay $\mathcal{B}$ is 7.6\%~\cite{Olive:2016xmw}. The uncertainty for $N_{\text{MB}}$ accounts for the inefficiency of the event selection and the trigger in selecting hadronic events~\cite{Khachatryan:2016odn}. The \TAA uncertainty is $+2.8\%, -3.4\%$~\cite{Khachatryan:2016odn}. In the calculations of the systematic uncertainties of the $\PBzs$ meson \RAA and the $\mathrm{\RAA}$ ratio between $\PBzs$ and $\PBp$, correlated uncertainties from the track and muon reconstruction and identification are partially canceled.

The values for each systematic uncertainty source are listed in Table~\ref{tab:sys_sum_Bu}.

\begin{table*}[h]
\begin{center}
\topcaption{Summary of systematic uncertainties in percentage (\%) from each source in \pp and \PbPb analyses.}
\label{tab:sys_sum_Bu}
  \begin{tabular}{ l  {c}@{\hspace*{5pt}} c  c  c {c}@{\hspace*{5pt}} c  c }
    Collision system && \multicolumn{3}{ c }{\pp} && \multicolumn{2}{ c }{\PbPb} \\
	\cline{1-1}\cline{3-5}\cline{7-8}
    \pt interval (\GeVcns) && [7,15] & [15,20] & [20,50] && [7,15] & [15,50] \\
    Signal modeling && 2.5 & 0.7 & 0.7 && 4.2 & 3.5 \\
    Background modeling && 3.4 & 1.6 & 1.6 && 8.7 & 0.68 \\
    \PBzs\ selection  && 15 & 2.6 & 2.6 && 19 & 8.6 \\
    \PBzs\ acceptance && 1.7 & 1.4 & 1.7 && 1.7 & 1.7 \\
    \PBzs\ efficiency && 6.5 & 0.5 & 0.9 && 7.9 & 3.8 \\
    MC sample size && 0.8 & 0.8 & 0.5 && 4.9 & 2.1  \\
    Muon trigger, reconstruction, and identification && 4.4 & 3.3 & 3.0 && 5.1 & 3.8 \\
    Hadron tracking efficiency && 8 & 8 & 8 && 12 & 12 \\
    Total && 19 & 9.4 & 9.3 && 26 & 16 \\
	[\cmsTabSkip]
    Branching fractions && \multicolumn{6}{ c }{7.6} \\
    Number of minimum bias events in PbPb data && \multicolumn{3}{ c }{\NA} && \multicolumn{2}{ c }{2} \\
    \TAA && \multicolumn{3}{ c }{\NA} && \multicolumn{2}{ c }{$+$2.8/$-$3.4} \\
    Integrated luminosity of \pp data && \multicolumn{3}{ c }{2.3} && \multicolumn{2}{ c }{\NA} \\
	\hline
\end{tabular}
\end{center}
\end{table*}

\section{Results}

\begin{figure}[tbhp]
\centering
\includegraphics[width=.45\textwidth]{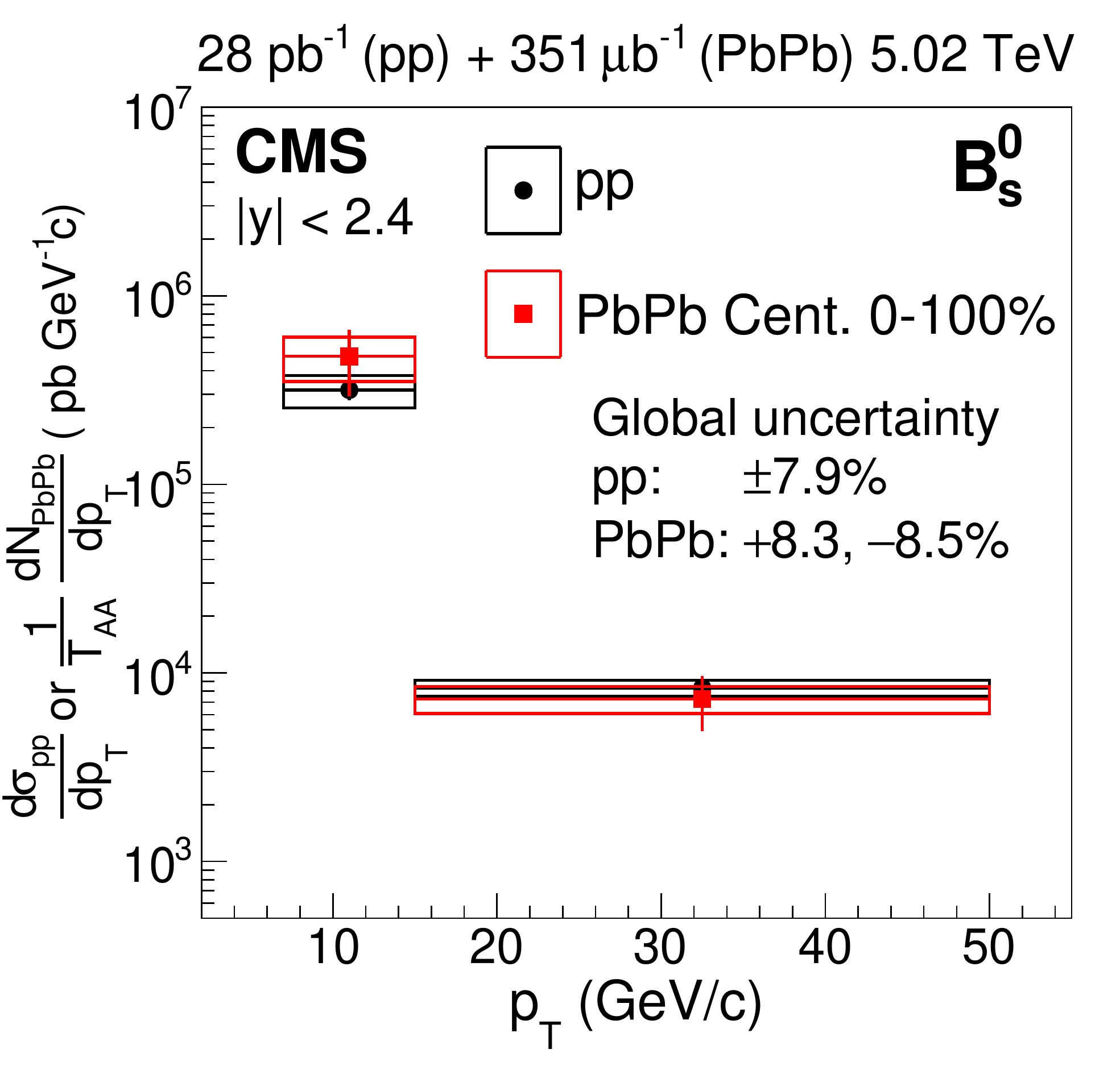}
\includegraphics[width=.45\textwidth]{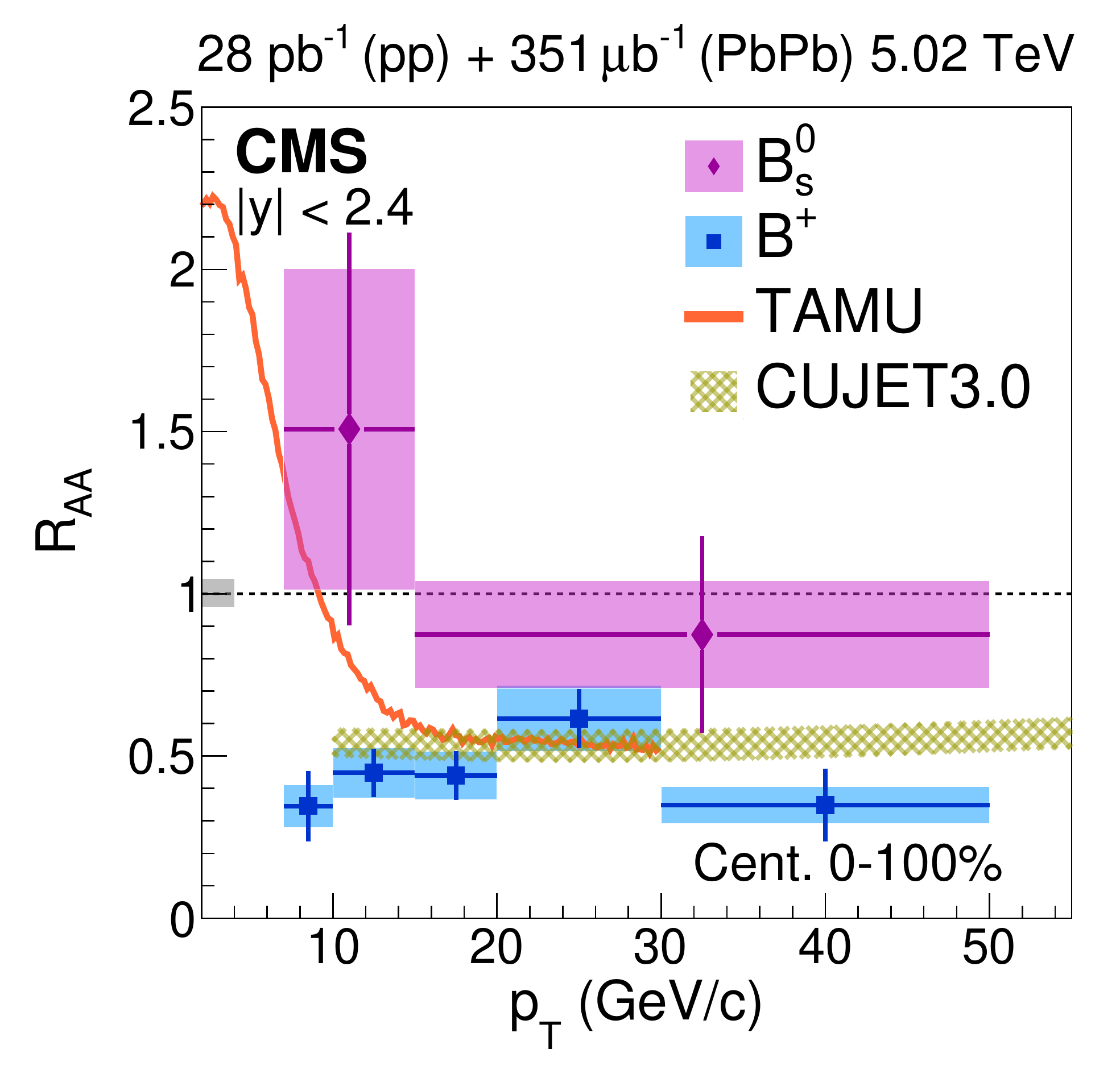}
\caption{(\cmsLeft) The \pt-differential production cross section of $\PBzs$ mesons in \pp collisions and the \pt-differential corrected yield of \PBzs\ mesons scaled by \TAA in \PbPb collisions at $\sqrtsNN=5.02\TeV$ in two \pt intervals from 7 to 50\GeVc.
The vertical bars (boxes) correspond to statistical (systematic) uncertainties. The global systematic uncertainty comprises the uncertainties in \TAA, $N_{\text{MB}}$, and $\mathcal{B}$.
(\cmsRight) The nuclear modification factor \RAA of $\PBzs$ measured in \PbPb collisions at $\sqrtsNN=5.02\TeV$ from 7 to 50\GeVc. The vertical bars (boxes) correspond to statistical (systematic) uncertainties. The \PBp\ \RAA measurement~\cite{Sirunyan:2017oug} is also shown for comparison. The global systematic uncertainty, represented by the grey  box at $\RAA=1$, comprises the uncertainties in the integrated luminosity measurement and \TAA value. Two $\PBzs$ theoretical calculations are also shown for comparison: TAMU~\cite{He:2011qa, He:2014cla} and CUJET3.0~\cite{Xu:2015bbz, Xu:2014tda, Xu:2014ica}. The line width of the theoretical calculation from Refs.~\cite{He:2011qa, He:2014cla} represents the size of its statistical uncertainty.}
\label{fig:raaall}
\end{figure}

In Fig.~\ref{fig:crosssections} and in the \cmsLeft panel of Fig.~\ref{fig:raaall}, the $\pt$-differential production cross sections in \pp and \PbPb collisions measured in the interval $\abs{y} < 2.4$ are presented. The \pp results are compared to the predictions of FONLL calculations~\cite{FONLLcharmbottomPP3}. The FONLL reference cross section is obtained by multiplying the FONLL total {\cPqb} quark production~\cite{FONLLcharmbottomPP1,FONLLcharmbottomPP2,FONLLcharmbottomPP3} by the world-average production fraction of $\PBzs$ of 10.3\%~\cite{Olive:2016xmw}. The $\PBzs$ FONLL prediction is consistent with the measured $\PBzs$ \pp spectrum within the uncertainties. The measured spectrum has a smaller uncertainty than that of the FONLL calculation. 

The nuclear modification factor \RAA, shown in Fig.~\ref{fig:raaall}, is computed as:
\begin{linenomath}
\begin{equation}
 \RAA (\pt) = \frac{1}{\TAA} \frac{{\rd}N^{\PBzs}_\PbPb}{{\rd}\pt} \bigg/ \frac{{\rd}\sigma^{\PBzs}_{\pp}}{{\rd}\pt}.
 \label{eq:RAA}
\end{equation}
\end{linenomath}
The \PBzs\ meson \RAA is $1.5 \pm 0.6\stat\pm 0.5\syst$ for 7--15\GeVc, and $0.87 \pm 0.30\stat\pm 0.17\syst$ for 15--50\GeVc, respectively. In the \cmsRight panel of Fig.~\ref{fig:raaall}, the \RAA of \PBp\ mesons from a previous measurement~\cite{Sirunyan:2017oug} is also shown. Compared to the \PBp\ mesons, there is an indication of an enhancement for \PBzs\ mesons, which would be the expectation in the presence of a contribution from beauty recombination with strange quarks in heavy ion collisions. However, the \PBzs\ \RAA values are compatible with unity and their large uncertainties do not exclude a significant suppression. The \pt dependence of \RAA is compared to the \PBzs\ prediction of a perturbative QCD based model that includes both collisional and radiative energy loss, (CUJET3.0)~\cite{Xu:2015bbz, Xu:2014tda, Xu:2014ica}, and a transport model based on a Langevin equation that includes collisional energy loss and heavy quark diffusion in the medium, (TAMU)~\cite{He:2011qa, He:2014cla}. The difference between the two models below $\pt\sim15$\GeV reflects the contribution from recombination processes, which are included in the TAMU but not in the CUJET3.0 model. The results measured for $\pt>7$\GeVc have the power to disentangle the two models, albeit after an increase in precision, which can be achieved with a bigger data sample.

\begin{figure}[tbhp]
\centering
\includegraphics[width=.45\textwidth]{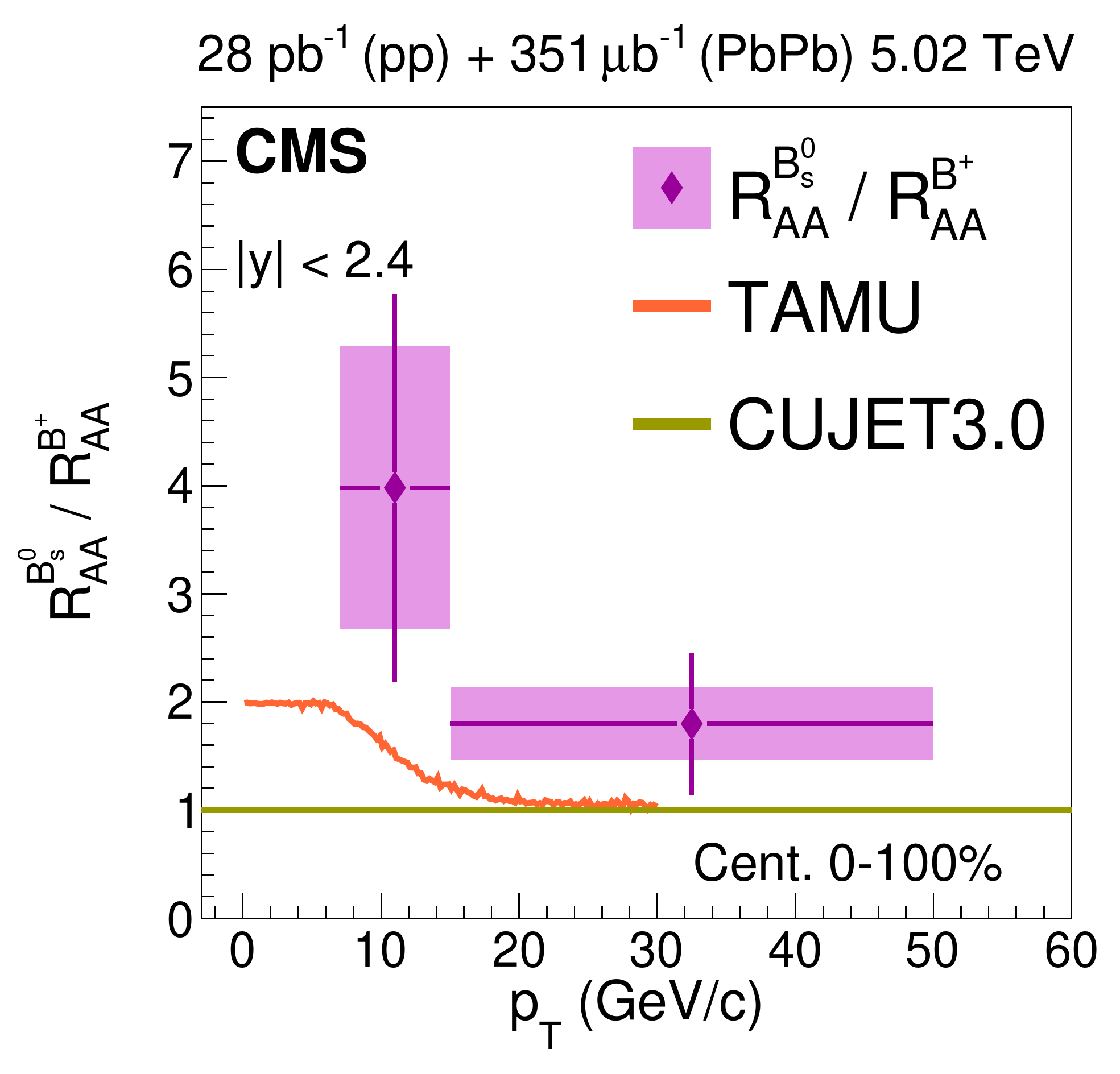}
\caption{The nuclear modification factor $\mathrm{\RAA}$ ratio between $\PBzs$ and $\PBp$ measured in \PbPb collisions at $\sqrtsNN=5.02\TeV$ from 7 to 50\GeVc. Two $\PBzs$ theoretical calculations are also shown for comparison: TAMU~\cite{He:2011qa, He:2014cla}, and CUJET3.0~\cite{Xu:2015bbz, Xu:2014tda, Xu:2014ica}.
}
\label{fig:raaratio}
\end{figure}

To further quantify the significance of a possible enhancement of the $\PBzs$/$\PBp$ ratio in \PbPb with respect to \pp collisions, the ratio between the $\PBzs$ and the \PBp\ \RAA is also calculated, canceling the systematic uncertainty sources that are common to both measurements (acceptance, tracking efficiency, and muon-related). The $\PBp$~$\RAA$ with a wider \pt binning (15--50\GeVc) is obtained by a $\PBp$ yield weighted average of the results from three \pt bins (15--20, 20--30 and 30--50\GeVc) presented in previous work~\cite{Sirunyan:2017oug}. The result is shown in Fig.~\ref{fig:raaratio}. The ratio is $4.0 \pm 1.8\stat\pm 1.3\syst$ for 7--15\GeVc, and $1.8 \pm 0.7\stat\pm 0.3\syst$ for 15--50\GeVc, respectively.  Assuming a Gaussian distribution with mean and width equal to that of the $\RAA$ ratio and its uncertainty (including statistical and systematic components added in quadrature), the hypothesis of the ratio values being consistent with unity (no enhancement) is tested with a $\chi^2$ test. The resulting p-values are 18\% and 28\% for 7--15 and 15--50\GeVc, respectively. This shows that, with a $p$-value cutoff of 5\%, the scenario of no enhancement cannot be rejected. This analysis demonstrates the capability of performing a fully reconstructed $\PBzs$ measurement in \PbPb collisions with the CMS detector.

\section{Summary}

The first measurement of the differential production cross section of $\PBzs$ mesons (including both charge conjugates) in both \pp and \PbPb collisions at a center-of-mass energy of $5.02\TeV$ per nucleon pair is presented. The $\PBzs$ and $\PABzs$ mesons are studied with the CMS detector at the LHC in the rapidity range $\abs{y}<2.4$ via the reconstruction of one of their exclusive hadronic decay channels, $\Bzerosdecay$. The nuclear modification factor \RAA of $\PBzs$ is measured in the transverse momentum range from 7 to 50\GeVc, inclusively for 0--100\% event centrality. A hint of an enhancement of the $\PBzs$/$\PBp$ ratio in \PbPb with respect to \pp collisions is seen. More precise measurements of the $\PBzs$ and $\PBpm$ mesons \RAA with the upcoming high-luminosity LHC heavy ion runs could provide further constraints on the relevance of recombination, a marker of deconfined matter, for beauty hadron production, and unambiguous information about the mechanisms of beauty hadronization in heavy ion collisions.

\begin{acknowledgments}
We congratulate our colleagues in the CERN accelerator departments for the excellent performance of the LHC and thank the technical and administrative staffs at CERN and at other CMS institutes for their contributions to the success of the CMS effort. In addition, we gratefully acknowledge the computing centers and personnel of the Worldwide LHC Computing Grid for delivering so effectively the computing infrastructure essential to our analyses. Finally, we acknowledge the enduring support for the construction and operation of the LHC and the CMS detector provided by the following funding agencies: BMBWF and FWF (Austria); FNRS and FWO (Belgium); CNPq, CAPES, FAPERJ, FAPERGS, and FAPESP (Brazil); MES (Bulgaria); CERN; CAS, MoST, and NSFC (China); COLCIENCIAS (Colombia); MSES and CSF (Croatia); RPF (Cyprus); SENESCYT (Ecuador); MoER, ERC IUT, and ERDF (Estonia); Academy of Finland, MEC, and HIP (Finland); CEA and CNRS/IN2P3 (France); BMBF, DFG, and HGF (Germany); GSRT (Greece); NKFIA (Hungary); DAE and DST (India); IPM (Iran); SFI (Ireland); INFN (Italy); MSIP and NRF (Republic of Korea); MES (Latvia); LAS (Lithuania); MOE and UM (Malaysia); BUAP, CINVESTAV, CONACYT, LNS, SEP, and UASLP-FAI (Mexico); MOS (Montenegro); MBIE (New Zealand); PAEC (Pakistan); MSHE and NSC (Poland); FCT (Portugal); JINR (Dubna); MON, RosAtom, RAS, RFBR, and NRC KI (Russia); MESTD (Serbia); SEIDI, CPAN, PCTI, and FEDER (Spain); MOSTR (Sri Lanka); Swiss Funding Agencies (Switzerland); MST (Taipei); ThEPCenter, IPST, STAR, and NSTDA (Thailand); TUBITAK and TAEK (Turkey); NASU and SFFR (Ukraine); STFC (United Kingdom); DOE and NSF (USA).

\hyphenation{Rachada-pisek} Individuals have received support from the Marie-Curie programme and the European Research Council and Horizon 2020 Grant, contract No. 675440 (European Union); the Leventis Foundation; the A. P. Sloan Foundation; the Alexander von Humboldt Foundation; the Belgian Federal Science Policy Office; the Fonds pour la Formation \`a la Recherche dans l'Industrie et dans l'Agriculture (FRIA-Belgium); the Agentschap voor Innovatie door Wetenschap en Technologie (IWT-Belgium); the F.R.S.-FNRS and FWO (Belgium) under the ``Excellence of Science - EOS" - be.h project n. 30820817; the Ministry of Education, Youth and Sports (MEYS) of the Czech Republic; the Lend\"ulet (``Momentum") Programme and the J\'anos Bolyai Research Scholarship of the Hungarian Academy of Sciences, the New National Excellence Program \'UNKP, the NKFIA research grants 123842, 123959, 124845, 124850 and 125105 (Hungary); the Council of Science and Industrial Research, India; the HOMING PLUS programme of the Foundation for Polish Science, cofinanced from European Union, Regional Development Fund, the Mobility Plus programme of the Ministry of Science and Higher Education, the National Science Center (Poland), contracts Harmonia 2014/14/M/ST2/00428, Opus 2014/13/B/ST2/02543, 2014/15/B/ST2/03998, and 2015/19/B/ST2/02861, Sonata-bis 2012/07/E/ST2/01406; the National Priorities Research Program by Qatar National Research Fund; the Programa Estatal de Fomento de la Investigaci{\'o}n Cient{\'i}fica y T{\'e}cnica de Excelencia Mar\'{\i}a de Maeztu, grant MDM-2015-0509 and the Programa Severo Ochoa del Principado de Asturias; the Thalis and Aristeia programmes cofinanced by EU-ESF and the Greek NSRF; the Rachadapisek Sompot Fund for Postdoctoral Fellowship, Chulalongkorn University and the Chulalongkorn Academic into Its 2nd Century Project Advancement Project (Thailand); the Welch Foundation, contract C-1845; and the Weston Havens Foundation (USA).
\end{acknowledgments}

\bibliography{auto_generated}

\providecommand{\href}[2]{#2}\begingroup\raggedright\begin{thebibliography}{10}%
\makeatletter
\providecommand{\hrefCMSnoop }[0]{\@secondoftwo}%
\makeatother
\providecommand{\doi}{\texttt{doi:}\begingroup \urlstyle{tt}\Url}

\bibitem{QGP1}
\href {http://www.jetp.ac.ru/cgi-bin/e/index/e/47/2/p212?a=list}{{\'E}.~V.
  Shuryak, ``Theory of hadronic plasma'',} \textit{ Sov. Phys. JETP} \textbf{
  47} (1978)
212.
%%CITATION = SPHJA,47,212;%%.

\bibitem{QGP2}
\hrefCMSnoop {}{J.~C. Collins and M.~J. Perry, ``Superdense matter: neutrons or
  asymptotically free quarks?'',} \textit{ Phys. Rev. Lett.} \textbf{ 34}
  (1975) 1353,
  \href{http://dx.doi.org/10.1103/PhysRevLett.34.1353}{\doi{10.1103/PhysRevLett.34.1353}}.

\bibitem{Karsch:2003jg}
\hrefCMSnoop {}{F.~Karsch and E.~Laermann, ``Thermodynamics and in-medium
  hadron properties from lattice {QCD}'',} in \textit{ Quark-Gluon Plasma III,
  R. Hwa (ed.), \textnormal{2003}}.
\newblock 2003.
\newblock
\href{http://www.arXiv.org/abs/hep-lat/0305025}{\texttt{arXiv:hep-lat/0305025}}.
\newblock
%%CITATION = HEP-LAT/0305025;%%.

\bibitem{Eloss1}
\href {http://lss.fnal.gov/archive/1982/pub/Pub-82-059-T.pdf}{J.~D. Bjorken,
  ``Energy loss of energetic partons in quark-gluon plasma: possible extinction
  of high $\pt$ jets in hadron-hadron collisions'',} Fermilab PUB 82-059-THY,
  1982.

\bibitem{Baier:2000mf}
\hrefCMSnoop {}{R.~Baier, D.~Schiff, and B.~G. Zakharov, ``Energy loss in
  perturbative {QCD}'',} \textit{ Ann. Rev. Nucl. Part. Sci.} \textbf{ 50}
  (2000) 37,
  \href{http://dx.doi.org/10.1146/annurev.nucl.50.1.37}{\doi{10.1146/annurev.nucl.50.1.37}},
\href{http://www.arXiv.org/abs/hep-ph/0002198}{\texttt{arXiv:hep-ph/0002198}}.
%%CITATION = HEP-PH/0002198;%%.

\bibitem{Chatrchyan:2011sx}
\hrefCMSnoop {}{{CMS Collaboration}, ``Observation and studies of jet quenching
  in {PbPb} collisions at {$\sqrtsNN =2.76 \TeV$}'',} \textit{ Phys. Rev. C}
  \textbf{ 84} (2011) 024906,
  \href{http://dx.doi.org/10.1103/PhysRevC.84.024906}{\doi{10.1103/PhysRevC.84.024906}},
\href{http://www.arXiv.org/abs/1102.1957}{\texttt{arXiv:1102.1957}}.
%%CITATION = ARXIV:1102.1957;%%.

\bibitem{Aad:2010bu}
\hrefCMSnoop {}{{ATLAS Collaboration}, ``Observation of a centrality-dependent
  dijet asymmetry in lead-lead collisions at {$\sqrtsNN =2.76 \TeV$} with the
  {ATLAS} detector at the {LHC}'',} \textit{ Phys. Rev. Lett.} \textbf{ 105}
  (2010) 252303,
  \href{http://dx.doi.org/10.1103/PhysRevLett.105.252303}{\doi{10.1103/PhysRevLett.105.252303}},
\href{http://www.arXiv.org/abs/1011.6182}{\texttt{arXiv:1011.6182}}.
%%CITATION = ARXIV:1011.6182;%%.

\bibitem{Andronic:2015wma}
\hrefCMSnoop {}{A.~Andronic {et~al.}, ``Heavy-flavour and quarkonium production
  in the {LHC} era: from proton-proton to heavy-ion collisions'',} \textit{
  Eur. Phys. J. C} \textbf{ 76} (2016) 107,
  \href{http://dx.doi.org/10.1140/epjc/s10052-015-3819-5}{\doi{10.1140/epjc/s10052-015-3819-5}},
\href{http://www.arXiv.org/abs/1506.03981}{\texttt{arXiv:1506.03981}}.
%%CITATION = ARXIV:1506.03981;%%.

\bibitem{PhysRevLett.48.1066}
\hrefCMSnoop {}{J.~Rafelski and B.~M{\"u}ller, ``{Strangeness Production in the
  Quark - Gluon Plasma}'',} \textit{ Phys. Rev. Lett.} \textbf{ 48} (1982)
  1066,
  \href{http://dx.doi.org/10.1103/PhysRevLett.48.1066}{\doi{10.1103/PhysRevLett.48.1066}}.
[Erratum: \DOI{10.1103/PhysRevLett.56.2334}].
%%CITATION = PRLTA,48,1066;%%.

\bibitem{Agakishiev:2011ar}
\hrefCMSnoop {}{{STAR} Collaboration, ``Strangeness enhancement in {Cu+Cu} and
  {Au+Au} collisions at {$\sqrtsNN = 200$ GeV}'',} \textit{ Phys. Rev. Lett.}
  \textbf{ 108} (2012) 072301,
  \href{http://dx.doi.org/10.1103/PhysRevLett.108.072301}{\doi{10.1103/PhysRevLett.108.072301}},
\href{http://www.arXiv.org/abs/1107.2955}{\texttt{arXiv:1107.2955}}.
%%CITATION = ARXIV:1107.2955;%%.

\bibitem{Abelev:2008zk}
\hrefCMSnoop {}{{STAR} Collaboration, ``Energy and system size dependence of
  phi meson production in {Cu+Cu} and {Au+Au} collisions'',} \textit{ Phys.
  Lett. B} \textbf{ 673} (2009) 183,
  \href{http://dx.doi.org/10.1016/j.physletb.2009.02.037}{\doi{10.1016/j.physletb.2009.02.037}},
\href{http://www.arXiv.org/abs/0810.4979}{\texttt{arXiv:0810.4979}}.
%%CITATION = ARXIV:0810.4979;%%.

\bibitem{Arsene:2009jg}
\hrefCMSnoop {}{{BRAHMS} Collaboration, ``Kaon and pion production in central
  {Au+Au} collisions at {$\sqrtsNN = 62.4$\GeV}'',} \textit{ Phys. Lett. B}
  \textbf{ 687} (2010) 36,
  \href{http://dx.doi.org/10.1016/j.physletb.2010.02.078}{\doi{10.1016/j.physletb.2010.02.078}},
\href{http://www.arXiv.org/abs/0911.2586}{\texttt{arXiv:0911.2586}}.
%%CITATION = ARXIV:0911.2586;%%.

\bibitem{Adamczyk:2017wsl}
\hrefCMSnoop {}{{STAR} Collaboration, ``Collision energy dependence of moments
  of net-kaon multiplicity distributions at {RHIC}'',} \textit{ Phys. Lett. B}
  \textbf{ 785} (2018) 551,
  \href{http://dx.doi.org/10.1016/j.physletb.2018.07.066}{\doi{10.1016/j.physletb.2018.07.066}},
\href{http://www.arXiv.org/abs/1709.00773}{\texttt{arXiv:1709.00773}}.
%%CITATION = ARXIV:1709.00773;%%.

\bibitem{Adare:2015ema}
\hrefCMSnoop {}{{PHENIX} Collaboration, ``{$\phi$ meson production in the
  forward/backward rapidity region in Cu$+$Au collisions at $\sqrtsNN=200$
  GeV}'',} \textit{ Phys. Rev. C} \textbf{ 93} (2016), no.~2, 024904,
  \href{http://dx.doi.org/10.1103/PhysRevC.93.024904}{\doi{10.1103/PhysRevC.93.024904}},
\href{http://www.arXiv.org/abs/1509.06337}{\texttt{arXiv:1509.06337}}.
%%CITATION = ARXIV:1509.06337;%%.

\bibitem{ALICE:2017jyt}
\hrefCMSnoop {}{{ALICE Collaboration}, ``Enhanced production of multi-strange
  hadrons in high-multiplicity proton-proton collisions'',} \textit{ Nature
  Phys.} \textbf{ 13} (2017) 535,
  \href{http://dx.doi.org/10.1038/nphys4111}{\doi{10.1038/nphys4111}},
\href{http://www.arXiv.org/abs/1606.07424}{\texttt{arXiv:1606.07424}}.
%%CITATION = ARXIV:1606.07424;%%.

\bibitem{Abelev:2013xaa}
\hrefCMSnoop {}{{ALICE Collaboration}, ``{$K^0_S$ and $\Lambda$ production in
  Pb-Pb collisions at $\sqrt{s_{NN}}$ = 2.76 TeV}'',} \textit{ Phys. Rev.
  Lett.} \textbf{ 111} (2013) 222301,
  \href{http://dx.doi.org/10.1103/PhysRevLett.111.222301}{\doi{10.1103/PhysRevLett.111.222301}},
\href{http://www.arXiv.org/abs/1307.5530}{\texttt{arXiv:1307.5530}}.
%%CITATION = ARXIV:1307.5530;%%.

\bibitem{He:2014cla}
\hrefCMSnoop {}{M.~He, R.~J. Fries, and R.~Rapp, ``Heavy flavor at the large
  hadron collider in a strong coupling approach'',} \textit{ Phys. Lett. B}
  \textbf{ 735} (2014) 445,
  \href{http://dx.doi.org/10.1016/j.physletb.2014.05.050}{\doi{10.1016/j.physletb.2014.05.050}},
\href{http://www.arXiv.org/abs/1401.3817}{\texttt{arXiv:1401.3817}}.
%%CITATION = ARXIV:1401.3817;%%.

\bibitem{Molnar:2003ff}
\hrefCMSnoop {}{D.~Molnar and S.~A. Voloshin, ``Elliptic flow at large
  transverse momenta from quark coalescence'',} \textit{ Phys. Rev. Lett.}
  \textbf{ 91} (2003) 092301,
  \href{http://dx.doi.org/10.1103/PhysRevLett.91.092301}{\doi{10.1103/PhysRevLett.91.092301}},
\href{http://www.arXiv.org/abs/nucl-th/0302014}{\texttt{arXiv:nucl-th/0302014}}.
%%CITATION = NUCL-TH/0302014;%%.

\bibitem{Greco:2003mm}
\hrefCMSnoop {}{V.~Greco, C.~M. Ko, and P.~Levai, ``Parton coalescence at
  {RHIC}'',} \textit{ Phys. Rev. C} \textbf{ 68} (2003) 034904,
  \href{http://dx.doi.org/10.1103/PhysRevC.68.034904}{\doi{10.1103/PhysRevC.68.034904}},
\href{http://www.arXiv.org/abs/nucl-th/0305024}{\texttt{arXiv:nucl-th/0305024}}.
%%CITATION = NUCL-TH/0305024;%%.

\bibitem{Greco:2003vf}
\hrefCMSnoop {}{V.~Greco, C.~M. Ko, and R.~Rapp, ``Quark coalescence for
  charmed mesons in ultrarelativistic heavy ion collisions'',} \textit{ Phys.
  Lett. B} \textbf{ 595} (2004) 202,
  \href{http://dx.doi.org/10.1016/j.physletb.2004.06.064}{\doi{10.1016/j.physletb.2004.06.064}},
\href{http://www.arXiv.org/abs/nucl-th/0312100}{\texttt{arXiv:nucl-th/0312100}}.
%%CITATION = NUCL-TH/0312100;%%.

\bibitem{Acharya:2018hre}
\hrefCMSnoop {}{{ALICE Collaboration}, ``{Measurement of D$^{0}$, D$^{+}$,
  D$^{*+}$ and D$_{s}^{+}$ production in Pb-Pb collisions at $
  \sqrtsNN=5.02$\TeV}'',} \textit{ JHEP} \textbf{ 10} (2018) 174,
  \href{http://dx.doi.org/10.1007/JHEP10(2018)174}{\doi{10.1007/JHEP10(2018)174}},
\href{http://www.arXiv.org/abs/1804.09083}{\texttt{arXiv:1804.09083}}.
%%CITATION = ARXIV:1804.09083;%%.

\bibitem{Chatrchyan:2011vh}
\hrefCMSnoop {}{{CMS Collaboration}, ``Measurement of the {$\PBzs$} production
  cross section with {$ \PBzs \rightarrow \PJGy \phi $} decays in \pp
  collisions at {$\sqrts = 7\TeV$}'',} \textit{ Phys. Rev. D} \textbf{ 84}
  (2011) 052008,
  \href{http://dx.doi.org/10.1103/PhysRevD.84.052008}{\doi{10.1103/PhysRevD.84.052008}},
\href{http://www.arXiv.org/abs/1106.4048}{\texttt{arXiv:1106.4048}}.
%%CITATION = ARXIV:1106.4048;%%.

\bibitem{Khachatryan:2015uja}
\hrefCMSnoop {}{{CMS Collaboration}, ``Study of {$B$} meson production in
  {p+Pb} collisions at {$\sqrtsNN = 5.02$\TeV} using exclusive hadronic
  decays'',} \textit{ Phys. Rev. Lett.} \textbf{ 116} (2016) 032301,
  \href{http://dx.doi.org/10.1103/PhysRevLett.116.032301}{\doi{10.1103/PhysRevLett.116.032301}},
\href{http://www.arXiv.org/abs/1508.06678}{\texttt{arXiv:1508.06678}}.
%%CITATION = ARXIV:1508.06678;%%.

\bibitem{FONLLcharmbottomPP1}
\hrefCMSnoop {}{M.~Cacciari, M.~Greco, and P.~Nason, ``The \pt spectrum in
  heavy-flavour hadroproduction'',} \textit{ JHEP} \textbf{ 05} (1998) 007,
  \href{http://dx.doi.org/10.1088/1126-6708/1998/05/007}{\doi{10.1088/1126-6708/1998/05/007}},
  \href{http://www.arXiv.org/abs/hep-ph/9803400}{\texttt{arXiv:hep-ph/9803400}}.

\bibitem{FONLLcharmbottomPP2}
\hrefCMSnoop {}{M.~Cacciari and P.~Nason, ``Charm cross sections for the
  {Tevatron Run II}'',} \textit{ JHEP} \textbf{ 09} (2003) 006,
  \href{http://dx.doi.org/10.1088/1126-6708/2003/09/006}{\doi{10.1088/1126-6708/2003/09/006}},
  \href{http://www.arXiv.org/abs/hep-ph/0306212}{\texttt{arXiv:hep-ph/0306212}}.

\bibitem{FONLLcharmbottomPP3}
M.~Cacciari\hrefCMSnoop {}{ {et~al.}, ``Theoretical predictions for charm and
  bottom production at the {LHC}'',} \textit{ JHEP} \textbf{ 10} (2012) 137,
  \href{http://dx.doi.org/10.1007/JHEP10(2012)137}{\doi{10.1007/JHEP10(2012)137}},
  \href{http://www.arXiv.org/abs/1205.6344}{\texttt{arXiv:1205.6344}}.

\bibitem{Sirunyan:2017oug}
\hrefCMSnoop {}{{CMS Collaboration}, ``Measurement of the {${B}^{\pm}$} meson
  nuclear modification factor in {Pb-Pb} collisions at
  {$\sqrtsNN=5.02$\TeV}'',} \textit{ Phys. Rev. Lett.} \textbf{ 119} (2017)
  152301,
  \href{http://dx.doi.org/10.1103/PhysRevLett.119.152301}{\doi{10.1103/PhysRevLett.119.152301}},
\href{http://www.arXiv.org/abs/1705.04727}{\texttt{arXiv:1705.04727}}.
%%CITATION = ARXIV:1705.04727;%%.

\bibitem{Olive:2016xmw}
\hrefCMSnoop {}{{Particle Data Group} Collaboration, ``Review of particle
  physics'',} \textit{ Chin. Phys. C} \textbf{ 40} (2016) 100001,
\href{http://dx.doi.org/10.1088/1674-1137/40/10/100001}{\doi{10.1088/1674-1137/40/10/100001}}.
%%CITATION = CHPHD,C40,100001;%%.

\bibitem{TRK-11-001}
\hrefCMSnoop {}{{CMS Collaboration}, ``Description and performance of track and
  primary-vertex reconstruction with the {CMS} tracker'',} \textit{ JINST}
  \textbf{ 9} (2014) P10009,
  \href{http://dx.doi.org/10.1088/1748-0221/9/10/P10009}{\doi{10.1088/1748-0221/9/10/P10009}},
\href{http://www.arXiv.org/abs/1405.6569}{\texttt{arXiv:1405.6569}}.
%%CITATION = ARXIV:1405.6569;%%.

\bibitem{Chatrchyan:2012xi}
\hrefCMSnoop {}{{CMS Collaboration}, ``Performance of {CMS} muon reconstruction
  in pp collision events at {$\sqrt{s}=7$\TeV}'',} \textit{ JINST} \textbf{ 7}
  (2012) P10002,
  \href{http://dx.doi.org/10.1088/1748-0221/7/10/P10002}{\doi{10.1088/1748-0221/7/10/P10002}},
\href{http://www.arXiv.org/abs/1206.4071}{\texttt{arXiv:1206.4071}}.
%%CITATION = ARXIV:1206.4071;%%.

\bibitem{bib_CMS}
\hrefCMSnoop {}{{CMS Collaboration}, ``The {CMS} experiment at the {CERN}
  {LHC}'',} \textit{ JINST} \textbf{ 3} (2008) S08004,
  \href{http://dx.doi.org/10.1088/1748-0221/3/08/S08004}{\doi{10.1088/1748-0221/3/08/S08004}}.

\bibitem{Sjostrand:2014zea}
T.~Sj{\"o}strand\hrefCMSnoop {}{ {et~al.}, ``An introduction to {PYTHIA
  8.2}'',} \textit{ Comput. Phys. Commun.} \textbf{ 191} (2015) 159,
  \href{http://dx.doi.org/10.1016/j.cpc.2015.01.024}{\doi{10.1016/j.cpc.2015.01.024}},
\href{http://www.arXiv.org/abs/1410.3012}{\texttt{arXiv:1410.3012}}.
%%CITATION = ARXIV:1410.3012;%%.

\bibitem{Khachatryan:2015pea}
\hrefCMSnoop {}{{CMS Collaboration}, ``Event generator tunes obtained from
  underlying event and multiparton scattering measurements'',} \textit{ Eur.
  Phys. J. C} \textbf{ 76} (2016)
  \href{http://dx.doi.org/10.1140/epjc/s10052-016-3988-x}{\doi{10.1140/epjc/s10052-016-3988-x}},
\href{http://www.arXiv.org/abs/1512.00815}{\texttt{arXiv:1512.00815}}.
%%CITATION = ARXIV:1512.00815;%%.

\bibitem{Allison:2016lfl}
\hrefCMSnoop {}{J.~Allison {et~al.}, ``Recent developments in {Geant4}'',}
  \textit{ Nucl. Instrum. Meth. A} \textbf{ 835} (2016) 186,
\href{http://dx.doi.org/10.1016/j.nima.2016.06.125}{\doi{10.1016/j.nima.2016.06.125}}.
%%CITATION = NUIMA,A835,186;%%.

\bibitem{Lange:2001uf}
\hrefCMSnoop {}{D.~J. Lange, ``The {EvtGen} particle decay simulation
  package'',} \textit{ Nucl. Instrum. Meth. A} \textbf{ 462} (2001) 152,
\href{http://dx.doi.org/10.1016/S0168-9002(01)00089-4}{\doi{10.1016/S0168-9002(01)00089-4}}.
%%CITATION = NUIMA,A462,152;%%.

\bibitem{Barberio:1990ms}
\hrefCMSnoop {}{E.~Barberio, B.~van Eijk, and Z.~Was, ``{PHOTOS}: A universal
  monte carlo for {QED} radiative corrections in decays'',} \textit{ Comput.
  Phys. Commun.} \textbf{ 66} (1991) 115,
\href{http://dx.doi.org/10.1016/0010-4655(91)90012-A}{\doi{10.1016/0010-4655(91)90012-A}}.
%%CITATION = CPHCB,66,115;%%.

\bibitem{Lokhtin:2005px}
\hrefCMSnoop {}{I.~P. Lokhtin and A.~M. Snigirev, ``A model of jet quenching in
  ultrarelativistic heavy ion collisions and high-{\pt} hadron spectra at
  {RHIC}'',} \textit{ Eur. Phys. J. C} \textbf{ 45} (2006) 211,
  \href{http://dx.doi.org/10.1140/epjc/s2005-02426-3}{\doi{10.1140/epjc/s2005-02426-3}},
\href{http://www.arXiv.org/abs/hep-ph/0506189}{\texttt{arXiv:hep-ph/0506189}}.
%%CITATION = HEP-PH/0506189;%%.

\bibitem{Khachatryan:2016odn}
\hrefCMSnoop {}{{CMS Collaboration}, ``Charged-particle nuclear modification
  factors in {PbPb} and {pPb} collisions at {$\sqrtsNN = 5.02$\TeV}'',}
  \textit{ JHEP} \textbf{ 04} (2017) 039,
  \href{http://dx.doi.org/10.1007/JHEP04(2017)039}{\doi{10.1007/JHEP04(2017)039}},
\href{http://www.arXiv.org/abs/1611.01664}{\texttt{arXiv:1611.01664}}.
%%CITATION = ARXIV:1611.01664;%%.

\bibitem{Khachatryan:2010xs}
\hrefCMSnoop {}{{CMS Collaboration}, ``Transverse momentum and pseudorapidity
  distributions of charged hadrons in pp collisions at {$\sqrts = 0.9$ and
  2.76\TeV}'',} \textit{ JHEP} \textbf{ 02} (2010) 041,
  \href{http://dx.doi.org/10.1007/JHEP02(2010)041}{\doi{10.1007/JHEP02(2010)041}},
\href{http://www.arXiv.org/abs/1002.0621}{\texttt{arXiv:1002.0621}}.
%%CITATION = ARXIV:1002.0621;%%.

\bibitem{CMS-PAS-LUM-16-001}
\href {https://cds.cern.ch/record/2235781}{{CMS Collaboration}, ``{CMS}
  luminosity calibration for the {\pp} reference run at
  {$\sqrt{s}=5.02\TeV$}'',} CMS Physics Analysis Summary CMS-PAS-LUM-16-001,
  2016.

\bibitem{PhysRevD.98.030001}
\hrefCMSnoop {}{{Particle Data Group}, M.~Tanabashi {et~al.}, ``Review of
  particle physics'',} \textit{ Phys. Rev. D} \textbf{ 98} (2018) 030001,
  \href{http://dx.doi.org/10.1103/PhysRevD.98.030001}{\doi{10.1103/PhysRevD.98.030001}}.

\bibitem{Hocker:2007ht}
\href {http://pos.sissa.it/archive/conferences/050/040/ACAT_040.pdf}{H.~Voss,
  A.~H{\"o}cker, J.~Stelzer, and F.~Tegenfeldt, ``{TMVA} -- the toolkit for
  multivariate data analysis'',} in \textit{ XIth International Workshop on
  Advanced Computing and Analysis Techniques in Physics Research (ACAT)},
  p.~40.
\newblock 2009.
\newblock
\href{http://www.arXiv.org/abs/physics/0703039}{\texttt{arXiv:physics/0703039}}.
\newblock
%%CITATION = PHYSICS/0703039;%%.

\bibitem{Miller:2007ri}
\hrefCMSnoop {}{M.~L. Miller, K.~Reygers, S.~J. Sanders, and P.~Steinberg,
  ``Glauber modeling in high-energy nuclear collisions'',} \textit{ Ann. Rev.
  Nucl. Part. Sci.} \textbf{ 57} (2007) 205,
  \href{http://dx.doi.org/10.1146/annurev.nucl.57.090506.123020}{\doi{10.1146/annurev.nucl.57.090506.123020}},
  \href{http://www.arXiv.org/abs/nucl-ex/0701025}{\texttt{arXiv:nucl-ex/0701025}}.

\bibitem{Khachatryan:2010xn}
\hrefCMSnoop {}{{CMS Collaboration}, ``Measurements of inclusive {W} and {Z}
  cross sections in pp collisions at {$\sqrts =7 \TeV$}'',} \textit{ JHEP}
  \textbf{ 01} (2011) 080,
  \href{http://dx.doi.org/10.1007/JHEP01(2011)080}{\doi{10.1007/JHEP01(2011)080}},
\href{http://www.arXiv.org/abs/1012.2466}{\texttt{arXiv:1012.2466}}.
%%CITATION = ARXIV:1012.2466;%%.

\bibitem{Xu:2015bbz}
\hrefCMSnoop {}{J.~Xu, J.~Liao, and M.~Gyulassy, ``Bridging soft-hard transport
  properties of quark-gluon plasmas with {CUJET3.0}'',} \textit{ JHEP} \textbf{
  02} (2016) 169,
  \href{http://dx.doi.org/10.1007/JHEP02(2016)169}{\doi{10.1007/JHEP02(2016)169}},
\href{http://www.arXiv.org/abs/1508.00552}{\texttt{arXiv:1508.00552}}.
%%CITATION = ARXIV:1508.00552;%%.

\bibitem{Xu:2014tda}
\hrefCMSnoop {}{J.~Xu, J.~Liao, and M.~Gyulassy, ``Consistency of perfect
  fluidity and jet quenching in semi-quark-gluon monopole plasmas'',} \textit{
  Chin. Phys. Lett.} \textbf{ 32} (2015) 092501,
  \href{http://dx.doi.org/10.1088/0256-307X/32/9/092501}{\doi{10.1088/0256-307X/32/9/092501}},
\href{http://www.arXiv.org/abs/1411.3673}{\texttt{arXiv:1411.3673}}.
%%CITATION = ARXIV:1411.3673;%%.

\bibitem{Xu:2014ica}
\hrefCMSnoop {}{J.~Xu, A.~Buzzatti, and M.~Gyulassy, ``Azimuthal jet flavor
  tomography with {CUJET2.0} of nuclear collisions at {RHIC} and {LHC}'',}
  \textit{ JHEP} \textbf{ 08} (2014) 063,
  \href{http://dx.doi.org/10.1007/JHEP08(2014)063}{\doi{10.1007/JHEP08(2014)063}},
\href{http://www.arXiv.org/abs/1402.2956}{\texttt{arXiv:1402.2956}}.
%%CITATION = ARXIV:1402.2956;%%.

\bibitem{He:2011qa}
\hrefCMSnoop {}{M.~He, R.~J. Fries, and R.~Rapp, ``Heavy-quark diffusion and
  hadronization in quark-gluon plasma'',} \textit{ Phys. Rev. C} \textbf{ 86}
  (2012) 014903,
  \href{http://dx.doi.org/10.1103/PhysRevC.86.014903}{\doi{10.1103/PhysRevC.86.014903}},
\href{http://www.arXiv.org/abs/1106.6006}{\texttt{arXiv:1106.6006}}.
%%CITATION = ARXIV:1106.6006;%%.

\end{thebibliography}\endgroup

\cleardoublepage \appendix\section{The CMS Collaboration \label{app:collab}}\begin{sloppypar}\hyphenpenalty=5000\widowpenalty=500\clubpenalty=5000\vskip\cmsinstskip
\textbf{Yerevan Physics Institute, Yerevan, Armenia}\\*[0pt]
A.M.~Sirunyan, A.~Tumasyan
\vskip\cmsinstskip
\textbf{Institut f\"{u}r Hochenergiephysik, Wien, Austria}\\*[0pt]
W.~Adam, F.~Ambrogi, E.~Asilar, T.~Bergauer, J.~Brandstetter, M.~Dragicevic, J.~Er\"{o}, A.~Escalante~Del~Valle, M.~Flechl, R.~Fr\"{u}hwirth\cmsAuthorMark{1}, V.M.~Ghete, J.~Hrubec, M.~Jeitler\cmsAuthorMark{1}, N.~Krammer, I.~Kr\"{a}tschmer, D.~Liko, T.~Madlener, I.~Mikulec, N.~Rad, H.~Rohringer, J.~Schieck\cmsAuthorMark{1}, R.~Sch\"{o}fbeck, M.~Spanring, D.~Spitzbart, A.~Taurok, W.~Waltenberger, J.~Wittmann, C.-E.~Wulz\cmsAuthorMark{1}, M.~Zarucki
% ...
\vskip\cmsinstskip
% ...
1:  Also at Vienna University of Technology, Vienna, Austria\\
% ...
\end{sloppypar}
\end{document}